\documentclass[aps,pre,preprint,superscriptaddress]{revtex4}
\usepackage{graphicx}
\usepackage{times}
\begin{document}
\def\d{\delta}
\def\D{\Delta}
\def\s{\sigma}
\def\g{\gamma}
\def\e{\epsilon}
\title{Binding properties and evolution of homodimers in protein-protein
  interaction networks} 
\author{Iaroslav Ispolatov\footnote[2]
{Corresponding author,
E-MAIL slava@ariadnegenomics.com, FAX(240) 453-6208. }}
\affiliation{Ariadne Genomics Inc., 9700 Great Seneca Highway,
Suite 113, Rockville, Maryland 20850, USA}
\altaffiliation{Permanent address:
Departamento de Fisica, Universidad de Santiago de Chile,
Casilla 302, Correo 2, Santiago, Chile}
\author{Anton Yuryev}
\affiliation{Ariadne Genomics
Inc., 9700 Great Seneca Highway, Suite 113, Rockville, Maryland
20850, USA}
\author{Ilya Mazo}
\affiliation{Ariadne Genomics
Inc., 9700 Great Seneca Highway, Suite 113, Rockville, Maryland
20850, USA}
\author{Sergei Maslov\footnote[3]
{Corresponding author,
E-MAIL maslov@bnl.gov, FAX (631) 344-2918.}}
\affiliation{ 
Department of Physics,
Brookhaven National Laboratory, Upton, New
  York 11973, USA}
\date{\today}
\begin{abstract}

We demonstrate that Protein-Protein Interaction (PPI) networks in
several eucaryotic organisms contain significantly more
self-interacting proteins than expected if such homodimers randomly
appeared in the course of the evolution. We also show that on
average homodimers have twice as many interaction partners
than non-self-interacting proteins.
More specifically the likelihood of a protein to physically
interact with itself was found to be proportional to the total
number of its binding partners. These properties of dimers are
are in agreement with a phenomenological model in which individual
proteins differ from each other by the degree of their
``stickiness'' or general propensity towards interaction with
other proteins including oneself.
A duplication of self-interacting proteins creates a pair of
paralogous proteins interacting with each other.
We show that such pairs
occur more frequently than could be explained by pure chance alone.
Similar to homodimers, proteins involved in heterodimers with their paralogs
on average have twice as many interacting partners than
the rest of the network. The likelihood of a pair of
paralogous proteins to interact with each other was also shown
to decrease with their sequence similarity. This all points to the conclusion
that most of interactions between paralogs are inherited from ancestral
homodimeric 
proteins, rather than established de novo after the duplication.
We finally discuss possible implications of our empirical
observations from functional and evolutionary standpoints.

\end{abstract}
\maketitle

\section{Introduction}

Many functionally important proteins such as receptors (G-protein
coupled receptors (Milligan et al. 2003), tyrosine kinase
receptors (Ronnstrand 2004)), enzyme complexes (Marianayagam et al. 2004), 
ion channels
(Simon and Goodenough 1998) and transcriptional factors (Amoutzias et
al. 2004) are homo- 
or hetero-dimers. For example, almost 70\% of enzymes listed in
the Brenda database (http://www.brenda.uni-koeln.de/) can self-interact to
form dimers 
or higher-order oligomers. As another example, G-protein coupled receptors
(Milligan et al. 2003), chemokine  (Mellado et al. 2001), cytokine (Langer et
al. 2004), and 
tyrosine kinase receptor (Ronnstrand 2004) families all use
oligomerization as a step in the pathway activation in response to
an agonist (Marianayagam et al. 2004). The examples of multi-protein complexes
containing homodimers include proteasome (Bochtler et al. 1999),
ribosome (Matadeen et al. 1999), nucleosome (Bentley et al. 1984). The
function of most filamentous proteins of the cytoskeleton such as
actin, myosin, spectrin, tubulin, etc, relies on their
oligomerization or polymerization. The ability to self-interact
confers several  structural and functional advantages to proteins,
including improved stability (Hattori et al. 2003, Dunbar et al. 2004)
control over the accessibility and specificity of active sites
(Marianayagam et al. 2004), and increased structural complexity. In addition,
self-association can help to minimize genome size, while
maintaining the advantages of modular complex formation. Protein
assembly into heterodimers has the combinatorial effect of
producing multiple species with different affinity to its
substrates and other biophysical characteristics, giving the cell
an instrument for fine-tuning its regulatory responses. Even
bigger variety of complexes contain (or are formed by) the
interacting paralogs, such as spliceosome (Mura et al. 2001),
acting promoting complex Apr2/3, membrane receptors (Rubin and Yarden 2001),
and transcription factors (Amoutzias et al. 2004).

While many specific dimerizing proteins are well studied and their
biological and structural properties have been established, little
is known about an overall topological influence and high-level
statistical properties of dimer distribution in protein networks.
The protein networks have recently become a subject of extensive
research by biologists as well as by scientists from other fields
interested in networks and graphs (see, for example,
(Spirin and Mirny 2003, Amoutzias et al. 2004, Wagner 2003,
Maslov and Sneppen 2002, Wuchty et al. 2003, Kim et al. 2002). Among various
studied types of protein-protein networks, a binding, or physical
interaction networks have several appealing properties that make
them a popular research subject: they are undirected, Boolean, and
the most extensive ones, in principle spanning over all proteins
present in a given organism. Several universal features of the
binding networks are believed to be established fairly well.
Examples include an apparent broad (scale-free) degree
distribution (Wagner 2003 and references therein), suppression
interactions between high-degree (hub) proteins (Maslov and Sneppen 2002), a
higher than randomly expected number of tightly linked sub-graphs
or cliques (Spirin and Mirny 2003), and evolutionary conservation of such
tightly linked sub-graphs (Wuchty et al. 2003). In this paper we
describe a systematic empirical study of topological properties of
the physical interaction network properties in the neighborhood of
homodimers (self-interacting proteins) as well as heterodimers
formed by paralogous proteins.

\section{basic observations}

We have assembled and analyzed the protein-protein interaction
(binding) networks from four organisms: the baker's yeast {\it S.
cerevisiae}, the nematode worm {\it C. elegans}, the fruit fly, {\it D.
melanogaster},and the human {\it H. sapiens} (see Materials and
Methods for details). 
\begin{table}[h]
\begin{ruledtabular}
\begin{tabular}{cccccc}
species & $N_{\mathrm{total}}$ & $N_{\mathrm{PPI}} $ & $N_{\mathrm{dimer}}$ &
$\langle k \rangle$ & $\langle k \rangle_{\mathrm{dimer}}$ \\ 
\hline yeast & 6713 & 4876            & 179   & $6.6 \pm  0.2$ & $12.4 \pm
1.2$ \\ 
\hline worm & 22268 & 3137            & 89    & $3.3 \pm  0.1$ & $13.1 \pm
2.2$ \\ 
\hline fly & 26148 & 6962             & 160   & $5.9 \pm  0.1$ & $14.2 \pm
1.2$ \\ 
\hline human & 25000 -- 50000  & 5331 & 1045  & $5.7 \pm  0.1$ & $14.0 \pm
0.6$ \\ 
\end{tabular}
\end{ruledtabular}
\caption{\label{tab_1} Estimated total number of proteins
$N_{\mathrm{total}}$, number of proteins involved in the protein-protein
interaction networks $N_{\mathrm{PPI}}$, the number of dimers or
self-interacting proteins $N_{\mathrm{dimer}}$, the average network degree
(the number of neighbors)  $\langle k \rangle$ over all $N_{\mathrm{PPI}}$
, and the average degree $\langle k \rangle_{\mathrm{dimer}}$ of
self-interacting proteins. }
\end{table}
The most apparent observation that follows
from the network data (Table~I) is that the number of
self-interacting proteins in all four organisms is substantially
higher than one would expect purely by chance. Indeed, in a
network with $ N $ proteins (each having at least one
interaction), a straightforward estimate assuming equal affinity
to itself and other proteins, suggests that a protein with the
connectivity (degree) $k$ would have a probability to bind to
itself equal to $k/N $. The total number of dimers then will be
the sum of this expression over all proteins, which is the average
connectivity, $\sum_{i=1}^{N_{}} k_i/N_{} \equiv \langle k
\rangle$ . As Table~I shows, the actual number of dimers
is 25-200 times higher than expected based on this simple-minded
hypothesis.

The abundance of dimers in all species suggests that their
functional importance has been preserved through the evolution. In
support of this conclusion we note that self-interacting proteins
also have about twice as many interaction partners compared to
non-dimers (Table~I). Indeed, the number of interaction
partners of a protein was shown before to be positively correlated
with its probability to be essential for the survival of the cell
and to be conserved in the course of evolution (Wuchty et al. 2003).

Sometimes the ease with which proteins form self-interactions has
purely structural (as opposed to functional) origin explained e.g.
by the domain swapping model (Bennet et al. 1994) 
Indeed, in the fully folded state
the individual structural components of a protein are expected to
make multiple binding contacts with each other. A pair of
identical (or homologous) proteins then might be able to use the
same set of contacts to physically interact with each other if
they encounter each other in a partially unfolded state.

It is interesting to note that average degrees of dimers
are about equal to each other in all four organisms studied here.
Average degrees of all proteins in the network are also
quite close to each other (an anomalously low
$\langle k \rangle \simeq 3$ of the worm network is
explained in the Materials and Methods section).
At present it is unclear if this apparent similarity is just a
coincidence or has some deeper explanations. In any case, the
inter- and intra-species comparison of these networks with each
other indicate that the data for protein-protein interaction in any of these
organisms are far from saturation and a considerable number of new interactions
is expected to be added to these networks in the future.

\section{Linear scaling}
To better understand connectivity patterns of homodimers in the
protein interaction network,
we studied how the likelihood of a protein to interact with itself
$P_{\mathrm{dimer}}(k)$ depends on its overall number of binding partners
(degree) $k$. 
$P_{\mathrm{dimer}}(k)$ was obtained by dividing a properly binned degree
histogram 
of all homodimers by the degree histogram of all proteins in the
network. 
\begin{figure}[h]
\centering
\includegraphics*[width=6in]{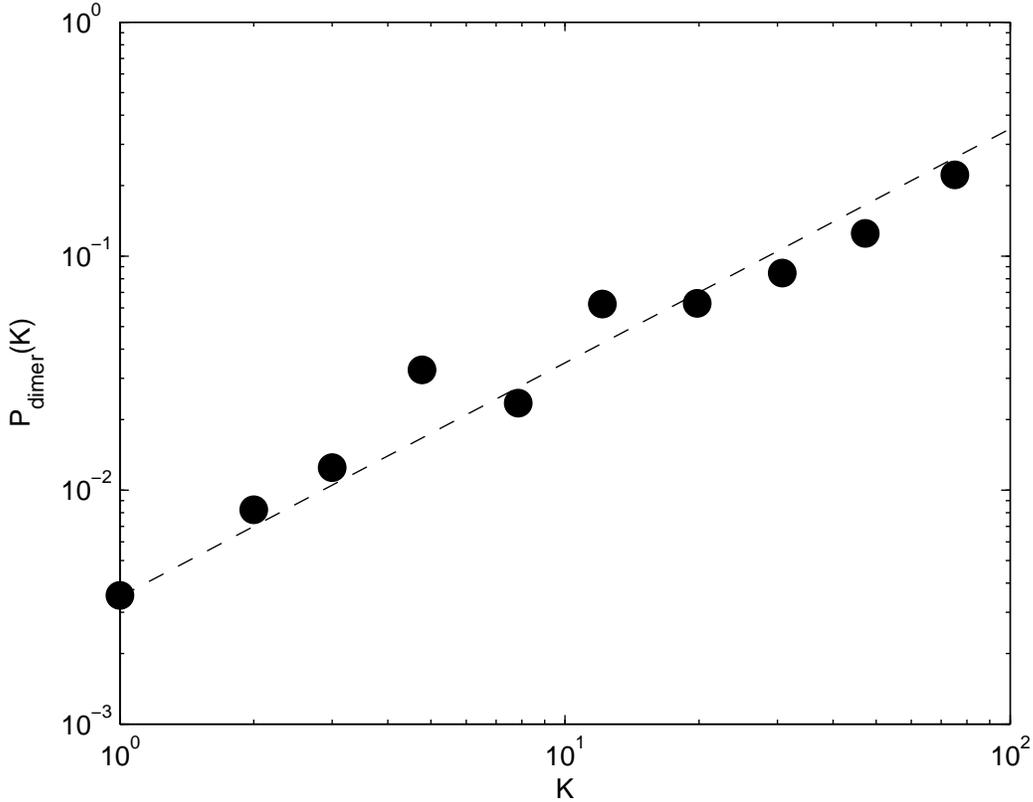}
\caption{The likelihood $P_{\mathrm{dimer}}(k)$ of a fly protein to
  self-interact 
plotted vs its degree $k$ in the PPI network. The dashed
line is the linear fit $P_{\mathrm{dimer}}(k)=0.0035 k$.}
\label{fig_pdim_fly}
\end{figure}

Fig.~1  shows $P_{\mathrm{dimer}}(k)$ vs $k$ measured in the fly
data based mainly on the species-wide two-hybrid dataset of (Giot et al. 2004).
As one can see, the probability of
self-interaction linearly increases with the degree in the protein
network (the dashed line on the log-log plot in Fig.~1
has slope 1). The proportionality coefficient of this
linear increase can be interpreted as the
probability $p_{\mathrm{self}} \simeq 3.5 \times 10^{-3}$
that a given edge of a physical interaction network starting at a
certain protein ends up connecting this node with itself. It is
approximately 25 times
larger than the probability $p_{\mathrm{others}}=1/7000 \simeq 1.4 \times
10^{-4}$ that it will instead connect with a randomly selected
other node among approximately 7000 proteins present in the fly
interaction dataset. This is consistent with a larger than
expected number of homodimers discussed above.

The observation that the likelihood of a protein to interact with
itself linearly increases with the total number of its interaction
(binding) partners (Fig.~1) contains an important
information about the general mechanisms of such interactions. We
conjecture that every protein $i$ can be characterized by a unique
intrinsic parameter that we would refer to as its ``stickiness''
$\sigma_{i}$. This parameter quantifies protein's overall
propensity towards forming physical interactions. We further
assume that both the probability of a protein to interact with
itself as well as its probability to interact with other proteins are
proportional to this stickiness (albeit with different
coefficients as we saw above) and thus should linearly depend on
each other.
This rather plausible conjecture of the existence of a ``universal
propensity towards interactions'' of individual proteins in an organism
thus explains both the linear scaling in
Fig.~1 and our original observation that
self-interacting proteins in several organisms tend to have higher
than average number of binding partners in the physical
interaction network (Table~I). Indeed, by considering 
the homodimers, we automatically pick proteins with higher than average
stickiness and thus end up with a subset of proteins characterized by
a higher than average number of binding partners $k$.

It is important to emphasize that the proposed ``stickiness'' of
a protein should not be interpreted literally, that is as the
ability of a protein to unspecifically bind other proteins.
In fact, all interactions in our datasets (with the exception of
false positives) come from specific functionally relevant bindings
between proteins.
Instead, one should view the ``stickiness'' as a complex quantitative
characteristic 
of a protein which has contributions from 
such properties as the number and nature of its constituent
domains, the hydrophobicity of its surface,
the number of copies of the protein per cell, the extent of its evolutionary
conservation, 
the overall level of a ``cooperativity'' of the functional task it is involved,
etc. In some of our datasets (e.g. human), which are based on a large number of
small-scale experiments instead of a single genome-wide assay, the
``stickiness'' of a protein may also correlate with its
overall popularity, i.e. the number of publications
it was studied in.
\begin{figure}[h]
\centering
\includegraphics*[width=6in]{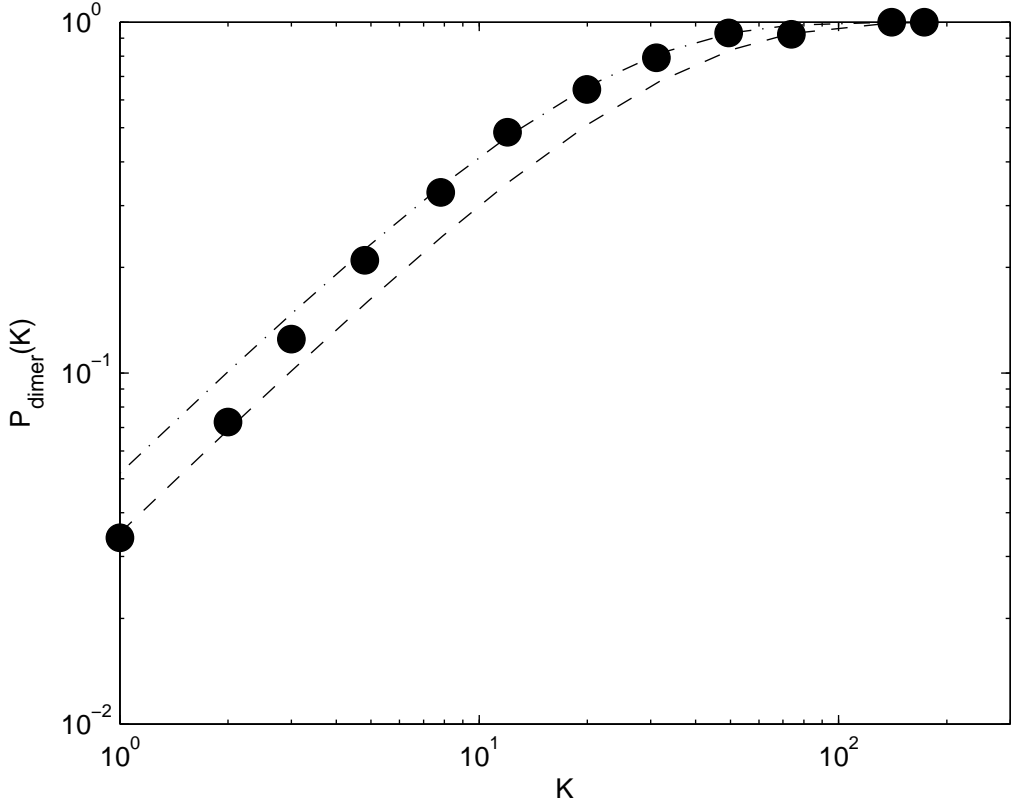}
\caption{The likelihood of a human protein to self-interact.
Dashed and dot-dashed lines are fits with the Eq.
(\ref{eq_saturation}) and $p_{\mathrm{self}}=0.035$ and
$p_{\mathrm{self}}=0.055$ 
correspondingly. The second value provides the best fit overall,
while the first value better fits the low $k$ region.}
\label{fig_pdim_human}
\end{figure}

Fig.~2 shows the
correlation between the propensity towards self-interactions and
the number of binding partners in the human dataset.
Here, as for the fly (see Fig.~1),
$P_{\mathrm{dimer}}(k)$ has a region of linear $k$-dependence. However, here
this 
region is limited  to small values of $k \lesssim 10$. For larger
values of $k$, $P_{\mathrm{dimer}}(k)$ starts to show saturation effects
and completely saturates at 1 for $k>100$. The saturation is
expected to follow a linear region as obviously no probability
could exceed 1. Moreover, it can be qualitatively described by the following
simple model. Suppose that each of the $k$ interaction links
starting at a given protein with a probability $p_{\mathrm{self}}$ ends at
the same protein, while with a probability $1-p_{\mathrm{self}}$ it selects
some other protein target. Then the chances that none of the $k$
links results in the formation of the homodimer are
$(1-p_{\mathrm{self}})^k$, while a homodimer is formed with a probability
\begin{equation}
P_{\mathrm{dimer}}(k)=1-(1-p_{\mathrm{self}})^k \qquad . \label{eq_saturation}
\end{equation}
For $k \ll 1/p_{\mathrm{self}}$ this expression yields a linear
$k$-dependence for $P_{\mathrm{dimer}}(k)$, as it was observed for the fly
data (Fig.~1). This general formula also fits
$P_{\mathrm{dimer}}(K)$ nicely over the whole range of $k$
(see dashed lines in the Fig.~2).

The fit with this formula provides an estimate of a propensity
towards self-interactions among human proteins: $p_{\mathrm{self}}^{(h)}
\simeq 0.03-0.06$ which is some $10$ times higher than in our fly
dataset. This is why the saturation of $P_{\mathrm{dimer}}(k)$ is clearly
visible in human but not in the fly. However, due to a vast
differences in the extent of coverage and sources of the data
describing protein-protein interactions in the human (interacting
protein pairs extracted from abstracts indexed in PubMed)
and the fly (a genome-wide two-hybrid
assay), different values of $p_{\mathrm{self}}$ do not have to reflect actual
differences between these two organisms.

\section{evolution of homodimers and interacting paralogs}
Interacting paralogous proteins (paralogous heterodimers) are
often thought (see, for example, Amoutzias et al. 2004) to be closely
related to the self-interacting proteins or homodimers.
Indeed, a duplication of a homodimer encoding gene in evolution results in an
appearance of a 
new pair (or several pairs for larger families)
of interacting paralogous proteins. Such interaction links
between paralogs could be destroyed with time as accumulation
of mutations in the constituent
proteins changes their three-dimensional shapes.
A binding between a pair of non-homodimeric
paralogous proteins may also appear {\it de novo} after
duplication event. Relative importance of these
two mechanisms of formation of paralogous heterodimers are
not universally agreed on (see e.g. Wagner 2003 for
a point of view favoring the {\it de novo} formation).

In this section we study pairs of interacting paralogs
present in our datasets. The purpose of this study is twofold:

Therefore the purpose of this section is twofold:

\begin{itemize}

\item  We first make a number of
empirical observations favoring the hereditary nature of
interactions between paralogs and confirming the relationship
between most of such heterodimers and their homodimeric ancestors.

\item  We then use a set of proteins interacting with their
paralogous partners to confirm and extend our empirical
observations about homodimers discussed in the previous section.
Due to an incomplete and noisy nature of essentially any data
describing genome-wide PPI networks there is only partial
overlap between sets of homodimers and interacting paralogs.
Thus the addition of interacting paralogs to the set of homodimers
allows us to considerably improve the statistics of our analysis.

\end{itemize}

We first just count the number of linked paralogous
pairs $n_{\mathrm{linked \ paralogs}}$ in each data set. If most links between
paralogs were 
indeed inherited from homodimeric ancestors, $n_{\mathrm{linked \ paralogs}}$
should 
be significantly higher than $n_{\mathrm{linked \ random}}$ the number of
links one expects to 
find between the same number $N_{\mathrm{paralogous \ pairs}}$ of randomly
selected pairs of 
non-paralogous proteins. Indeed, as we demonstrated in the
previous sections all four organisms included in our study are
characterized by an unusually large number of homodimers.
If on the other hand most links between
paralogous proteins were established {\it de novo} after the
duplication there is no reason to expect the number of such
links to be unusually large compared to a random set of protein
pairs. 
\begin{table}[h]
\begin{ruledtabular}
\begin{tabular}{cccccc}
species & $N_{\mathrm{paralogous \ pairs}}$ & $n_{\mathrm{linked \ paralogs}}$
& $n_{\mathrm{linked \ random}}$ & $\langle k \rangle_{\mathrm{linked \
    paralogs}} $ & $\langle k \rangle_{\mathrm{dimer}}$ \\ 
\hline yeast & 3409  & 251 & $4 \pm  2$ & $14.3 \pm 1.9 $ & $12.4 \pm  1.2$\\
\hline fly   & 12991 & 142 & $11\pm  3$ & $11.1 \pm  1.0$ &  $14.2 \pm 1.2$\\
\hline worm  & 3480  & 105 & $3 \pm  2$ & $ 5.8 \pm  0.9$ &  $13.1 \pm 2.2$\\
\hline human & 21562 &1280 & $24\pm  5$ & $10.2 \pm  0.6$ & $14.0 \pm  0.6$ \\
\end{tabular}
\end{ruledtabular}
\caption{\label{tab_2} The number of linked pairs of paralogous proteins
$n_{\mathrm{linked \ paralogs}}$,
the number of linked pairs $n_{\mathrm{linked \ paralogs}}$ expected by pure
chance alone, 
the average degree $\langle k \rangle_{\mathrm{linked \ paralogs}}$
of proteins known to interact
with some of their paralogs , and the
average degree $\langle k \rangle_{\mathrm{dimer}}$ of
self-interacting (dimer) proteins.}
\end{table}
The results presented in Table~II strongly support
the hereditary origin of most paralogous heterodimers: for all species
$n_{\mathrm{linked \ paralogs}}$ is much larger than  $n_{\mathrm{linked \
    random}}$ 
(by several orders of magnitude.)
This a strong evidence for the hereditary rather
than the {\it de novo} origin of the paralog-paralog links.
\begin{figure}[h]
\centering
\includegraphics*[width=5.5in]{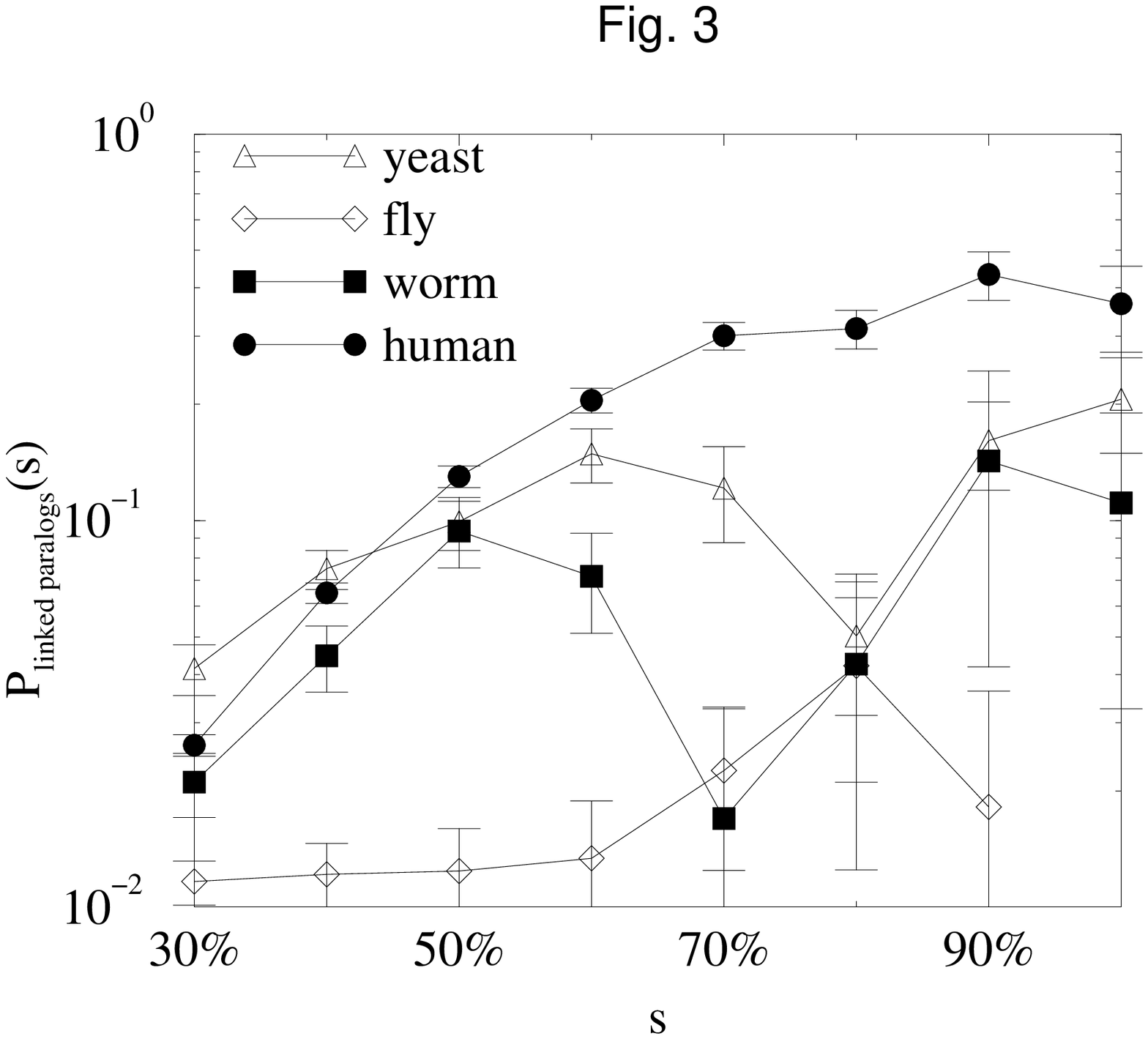}
\caption{\label{fig_dos} The probability for two paralogous proteins
to bind to each other $P_{\mathrm{linked \ paralogs}}$ vs their
sequence similarity $s$ for
(top to the bottom) human, yeast, worm, and fly.
Even the most distant paralogs are more likely to interact with each other
than a randomly selected pair of proteins. Such randomly expected probability
is 
equal to
$1.1 \times 10^{-3}$ in the human,
$1.3 \times 10^{-3}$  in the yeast,
$1.1 \times 10^{-3}$  in the worm, and
$0.8 \times 10^{-3}$ in the fly dataset.
}
\end{figure}
Another strong argument for the hereditary hypothesis follows from
Fig.~3. This figure reveals that the further paralogs
diverge in their amino-acid sequences, the smaller is the probability of them
to 
be linked to each other. This suggests
that typically pairs of linked paralogs gradually loose inherited interactions
rather than establish new ones.

Thus we conclude that most interacting paralogs present in our
data were created by duplication of homodimeric proteins.
A final argument in support of this conclusion is that
the average number of binding partners of interacting paralogs
$\langle k \rangle_{\mathrm{linked \ paralogs}}$ is indistinguishable from
that of homodimers $\langle k \rangle_{\mathrm{dimer}}$ and is
some $2-3$ times higher than the average over the whole
network (see Tables~I,II).

\begin{table}[h]
\begin{ruledtabular}
\begin{tabular}{ccccccc}
species & $N_{\mathrm{PPI}}$ & $N_{\mathrm{PPI-p}}$ & $N_{\mathrm{l-p}}$ &
$N_{\mathrm{dimer}}$ & $N_{\mathrm{d-p}}$ & $N_{\mathrm{d-l-p}}$ \\ 
\hline                                                          
yeast & 4876 &  1682 &    321 &     179 &      67 &   36 \\    
\hline                                                          
worm &  3137 &  1578 &    143 &      89 &      47 &   13 \\    
\hline                                                          
fly &   6962 &  2951 &    169 &     160 &      59 &   17 \\   
\hline                                                          
human & 5331 &  3840 &   1548 &    1045 &     789 &  460 \\    
\end{tabular}
\end{ruledtabular}
\caption{\label{tab_dos}
Numbers of certain types of proteins
for yeast, worm, fly, and
human: $N_{\mathrm{PPI}}$ - proteins present in the network,
$N_{\mathrm{PPI-p}}$ - network proteins with at least one paralog present in
the network 
$N_{\mathrm{l-p}}$ - proteins linked to at least one of their paralogs,
$N_{\mathrm{dimer}}$ - homodimers,
$N_{\mathrm{d-p}}$ - homodimers that have
at least one paralog among network proteins,
$N_{\mathrm{d-l-p}}$  -
homodimers linked to at least one of their paralogs.
}
\end{table}
Given that most paralogous heterodimers were at some point
formed from homodimers, one might assume that most proteins
involved in such heterodimeric complexes are homodimers.
However, it is far from being the case (see Table~III).
Such discrepancy is caused by two reasons, one
purely evolutionary while another  anthropogenic.
\begin{itemize}
\item As a result of substitutions in its amino-acid sequence
any protein might loose its ability to interact with
its paralog or to homodimerize. From Fig.~2 one can see
that many ancient duplicates of homodimers have lost links to their
ancestors.

\item The experimental data are far from being complete and many links,
  including 
self-interactions,  are simply not registered. The comparison between
sets of homodimers and interacting paralogs
may in principle be used to crudely estimate the completeness of our
knowledge of a protein network in a given organism.
\end{itemize}

\section{discussion}
Above we demonstrated
that self-interacting proteins tend to have connectivity
significantly above the average in the protein-protein interaction
network.  This phenomenon appears universally in
protein-protein interaction networks of all four model organisms
studied above. As a related phenomenon we found that interacting paralogs also
have increased connectivity, likely because most of them are
descendants of ancient self-interacting proteins.  We also have shown that
numbers of homodimers and interacting paralogs are both higher than
expected by pure chance alone. We unify
these phenomena by introducing a ``stickiness'' as a measure of
protein propensity for binding. Both the propensity of proteins
towards self-interactions and the degree of a protein in the
protein-protein interaction network are proportional to this parameter.
However, the dimerization probability apparently
has a larger proportionality coefficient. This is not very
surprising given a multitude of functional roles dimers (or
polymers) play in living cells. Dimerizing and oligomerizing
proteins are ubiquitous in all organisms and are present in the
most evolutionary conserved protein complexes (Marianayagam et al. 2004).

On the evolutionary side, we have confirmed that most links between
paralogs are most probably inherited from their dimerizing
ancestors. This does not exclude a possibility that some of these
links are formed after duplication as a result of random
mutations, but the relative number of such {\it de novo} created
links is relatively small. This conclusion has several implications for
the network topology. If a given
dimerizing protein has duplicated several times, it
leads to an appearance of a fully interconnected complex or clique of
paralogous heterodimers. In reality, some links inside this
complex are lost due the divergence of sequences of
paralogous proteins. Such loss of links may split a higher-order clique into
several lower-order ones or make it just a densely (yet not fully)
interconnected motif. A higher density of links
around dimers caused by  these remaining  heterodimeric links may
provide a qualitative explanation to the empirically observed abundance of
highly interconnected motifs and cliques in protein networks
(Spirin and Mirny 2003). Several simple models of network growth and
evolution due to gene duplications followed by subsequent
functional divergence of the resulting pair of paralogous proteins
lead to networks with an unrealistic bipartite topology in
which descendants of a particular protein never interact with
their paralogs (Kim et al. 2002).  Introduction of a large number of
heterodimers 
to the ancestral network in these models generates
frequent links between paralogs which
in the end gives rise to more realistic network topologies.

Finally, we would like to speculate on a general role that the
highly connected self-interacting proteins might play in the cell.  A single
protein molecule can simultaneously bind only a limited number of partners, at
most equal to the number of its functional domains. On the other
hand, most biological processes require many different proteins in
numbers far greater than the binding capacity of a single protein
molecule.  The protein components of large signaling or
biochemical pathways do not form large stable complexes containing
all proteins simultaneously. Yet all the necessary molecules must
be in a physical proximity to each other to form a
functional module.  This contradiction poses a question: how so
many different proteins could co-localize in a cell to correctly perform
a physiological function? A possible solution to this question involves
highly connected self-interacting proteins serving
as self-organizing centers for co-localization
of the pathway components.  The self-interaction (oligomerization)
of such proteins might function as
a general mechanism for sensing protein concentration
(Marianayagam et al. 2004) 
Indeed, a random increase of a local concentration of monomers
leads to their oligomerization and subsequently to the increase in
the concentration of binding sites for other pathway
components, increasing in turn their effective concentration.

\section{materials and methods}
The protein interaction data for all four species were obtained
from the Biological Association Network databases available from
Ariadne Genomics (http://www.ariadnegenomics.com/). The database for {\it
  H. sapiens} was 
derived from the Ariadne Genomics ResNet database, constructed from
the various literature sources using Medscan.  Medscan is the
Ariadne Genomics' proprietary natural language processing
technology (Novichkova et al. 2003, Daraselia et al. 2003). 
The databases for the
baker's yeast {\it S. cerevisiae}, the nematode worm {\it C. elegans}, and
the fruit fly, {\it D. melanogaster} were constructed by combining
the data from published high-throughput experiments with the
literature data obtained using Medscan technology.  For more
details on the construction of these databases please refer to the
PathwayAssist manual (http://www.ariadnegenomics.com/products/pathway.html).

Most of the PPI interactions
among fly proteins (20496 out of 20595 or 99.5\%) are extracted
from a single system-wide two-hybrid study (Giot et al. 2003), while most
of worm interactions (5286 out of 5309 or 99.5\%) are from a
large-scale two-hybrid study (Li et al. 2004). An abnormally small
average degree in the worm PPI network compared to
that of other organisms might be explained by the fact that,
unlike in the yeast (Ito et al. 2001) and the fly (Giot et al. 2003) cases, the
high-throughput two-hybrid
assay of worm proteins was not truly genome-wide.
Indeed, in (Li et al. 2004) the authors experimentally investigated 
interactions
of only 1873 baits (out of some 22000 worm proteins)
against genome-wide libraries of preys. This
resulted in an identification of 4027 distinct pairs of interacting
proteins which were subsequently extended to include a certain
number of in-silico predicted ``interologs''.
The average degree of these tested 1873 baits (or rather 632 of them
that we found among our network proteins) is approximately equal to 5.4.
Not only it is much higher than the average degree 3.3 reported for
all worm network proteins, but it is also remarkably close to the
$5.7-6.6$ range found in the other three organisms.

Lists of paralogous pairs and their sequence similarities
for all four species studied here were obtained by the
following procedure.
Amino-acid sequences of individual proteins
were obtained from the RefSeq database (http://www.ncbi.nlm.nih.gov/RefSeq/).
For each organism,
the sequences
were compared against themselves using the
BLASTp program with the expectation value cutoff equal to 0.001 (Altschul et
al. 1990). A
global alignment similarity 
was then computed by adding together numbers of similar
amino-acids from all non-overlapping locally aligned segments
and dividing this number by the geometric average of two protein
lengths. Thus gaps between the aligned segments were
considered to have zero similarity. In a case of overlapping segments
we took the one with the highest percent of similarity.
We estimated that about 2\% of the true homologs are not
recovered by this approach due to an incompleteness of the BLASTp
output for local alignment. Another sacrifice for quicker
calculation is an underestimation of the global alignment score by
5-10\% compared to more precise calculation after alignment using
the CLUSTALW  algorithm (Thompson et al. 1994).

To avoid including pairs of proteins similar over only one of their domains
we further restricted our set to only protein pairs with the similarity
higher than 30\%.
At the end all protein pairs that have been aligned by BLAST but omitted from
the final paralog list due to failing the similarity cutoff were searched for 
having
common paralogs.  If a common paralog was found, 
the pair was reinstated in the
paralog list, even though its similarity
is lower than the 30\%
cutoff.

\section{acknowledgment}
This work was supported by 1 R01 GM068954-01 grant from NIGMS.
Work at Brookhaven National Laboratory was carried out under
Contract No. DE-AC02-98CH10886, Division of Material Science, U.S.
Department of Energy.

\section{references}
Altschul, S.F., Gish, W., Miller, W., Myers, E.W., and Lipman,
                  D.J. 1990.  Basic local alignment search tool.
                  {\it J. Mol. Biol.} {\bf 215}: 403-410.

Amoutzias, G.D., Robertson, D.L., Oliver, S.G., and
  Bornberg-Bauer, E. 2004.  
  Convergent networks by single-gene duplications in higher eukaryotes.
  {\it EMBO Rep.} {\bf 5}: 274-279.

Bennett, M.J., Choe, S., and Eisenberg, D. 1994. Domain
  swapping: entangling alliances between proteins. {\it
  Proc. Natl. Acad. Sci. USA} {\bf 91}: 3127-3131.

Bentley, G.A., Lewit-Bentley, A., Finch, J.T., Podjarny, A.D., and Roth,
  M. 1984. Crystal structure of the nucleosome  core particle at 16 A
  resolution. {\it J Mol Biol.} {\bf 176}: 55-75.

Bochtler, M., Ditzel, L., Groll, M., Hartmann, C., and Huber, R. 1999. The
  proteasome. {\it Annu Rev Biophys Biomol Struct.} {\bf 28}: 295-317.

Daraselia, N.,  Yuryev, A., Egorov, S., Novichkova, S., Nikitin, A., 
                and Mazo, I. 2003. 
                  Extracting human protein interactions from MEDLINE using a
                  full-sentence parser. {\it Bioinformatics} {\bf 19}: 1-8.

Dunbar, A.Y., Kamada, Y., Jenkins, G.J., Lowe, E.R., Billecke, S.S., and
  Osawa, Y. 2004. 
  Ubiquitination and degradation of neuronal nitric-oxide synthase in
  vitro: dimer stabilization protects the enzyme from proteolysis. {\it Mol
  Pharmacol.} {\bf 66}:964-969.

Giot, L., Bader, J.S., Brouwer, C., Chaudhuri, A., Kuang, B., Li, Y., Hao,
Y.L., 
Ooi, C.E., Godwin, B. et al. 2003.
A protein interaction map of Drosophila melanogaster. {\it Science}
{\bf 302}: 1727-36.

Hattori, T., Ohoka, N., Inoue, Y., Hayashi, H., and Onozaki,
  K. 2003. C/EBP family transcription factors are degraded by the proteasome
  but stabilized by forming dimer. {\it Oncogene} {\bf 9}: 1273-1280.

Ito, T., Chiba, T., Ozawa, R., Yoshida, M., Hattori, M., Sakaki, Y. 2001. A
comprehensive two-hybrid analysis to explore the yeast protein
interactome. {\it Proc. Natl. Acad. Sci. USA} {\bf 98}: 4569-4574.

Kim, J, Krapivsky, P.L.,  Kahng, B., Redner, S. 2002. Infinite-order
                  percolation and giant fluctuations in a protein interaction
                  network. 
                  {\it Phys.\ Rev.\ E.} {\bf 66}: 055101-055105.

Langer, J.A., Cutrone, E.C., Kotenko, S. 2004. The Class II cytokine receptor
  (CRF2) family: overview and patterns of receptor-ligand interactions.
{\it Cytokine Growth Factor Rev.} {\bf 15}: 33-48.

Li, S., Armstrong, C.M., Bertin, N., Ge, H., Milstein, S.,
Boxem, M., Vidalain, P.O., Han, J.D., Chesneau, A., Hao, T. et al. 2004. 
A map of the interactome network of the metazoan C.
elegans. {\it Science} {\bf 303}: 540-543.

Maslov, S. and Sneppen, K. 2002. Specificity and stability in
                   topology of protein networks. {\it Science} {\bf 296}: 910.

Matadeen, R., Patwardhan, A., Gowen, B., Orlova, E.V., Pape, T., Cuff,
  M., Mueller, F., Brimacombe, R., and van Heel, M. 1999. The Escherichia coli
  large ribosomal subunit at 7.5 A resolution.
  {\it Structure Fold Des.} {\bf 12}: 1575-1583.

Marianayagam, N.J, Sunde, M., and Matthews, J.M. 2004.
The power of two: protein dimerization in biology. {\it Trends in
  Biochem. Sci.} {\bf 29}  618-625.

Mellado, M., Vila-Coro, A.J., Martinez, C., Rodriguez-Frade,
  J.M. 2001. Receptor dimerization: a key step in chemokine signaling.
{\it Cell Mol. Biol.} {\bf 47}: 575-582.

Milligan, G., Ramsay, D., Pascal, G., and Carrillo, J.J. 2003. 
GPCR dimerisation. {\it Life Sci.} {\bf 74}: 181-188.

Mura, C., Cascio, D., Sawaya, M.R., and Eisenberg, D.S. 2001. The crystal
  structure of a heptameric archaeal Sm protein: Implications for the
  eukaryotic snRNP core. {\it Proc. Natl. Acad. Sci. USA} {\bf 98}: 5532-5537.

Novichkova, S., Egorov, S., Daraselia, N. 2003. MedScan, a natural language
                  processing engine for MEDLINE 
                  abstracts. {\it Bioinformatics} {\bf 19}: 1699-1706.

Ronnstrand, L. 2004. Signal transduction via the stem cell factor
  receptor/c-Kit. {\it Cell Mol Life Sci.} {\bf 61}: 2535-2548.

Rubin, I., and Yarden, Y. 2001. The basic biology of HER2. {\it Ann Oncol.}
  {\bf 12} Suppl 1:S3-8.

Simon, A.M. and Goodenough, D.A., 1998.
Diverse functions of vertebrate gap junctions. {\it Trends Cell Biol.} 
{\bf 12}: 477-83.

Spirin, V., and Mirny, L.A. 2003.
  Protein complexes and functional modules in molecular
  networks. {\it Proc. Natl. Acad. Sci. USA} {\bf 100}: 12123-8.

Thompson, J.D., Higgins, D.G., and Gibson, T.J. 1994. CLUSTALW,
                  improving the sensitivity of progressive multiple
                  sequence alignment through sequence weighting,
                  position-specific gap penalties and weight matrix
                  choice. {\it Nucleic Acids Res.}  {\bf 22}: 4673-4680.

Wagner, A. 2003. How large protein interaction networks
                  evolve. {\it Proc. R. Soc. Lond. B} {\bf 270}: 457-466. 

Wuchty, S., Oltvai, Z.N., Barabasi, A.-L. 2003. Evolutionary
                   conservation of motif constituents within the yeast protein
                   interaction network. {\it Nature Genetics} {\bf 35}:
                   176-179.

\end{document}